\documentclass[twocolumn]{aastex63}

\usepackage{amsmath}
\usepackage{xspace}
\usepackage{layouts}
\usepackage{lineno}

\newcommand\Abacus{\textsc{Abacus}\xspace}

\newcommand{\hMsun}{\ensuremath{\mathit{h}^{-1}\ \mathrm{M_\odot}}}

\newcommand{\LCDM}{$\Lambda$CDM\xspace}

\newcommand{\bfr}{{\bf r}}

\newcommand{\bfF}{{\bf F}}

\shorttitle{Self-Similarity of Proper Softening}
\shortauthors{Garrison et al.}
\graphicspath{{./}{figures/}}

\begin{document}

\title{Good and Proper: Self-similarity of $N$-body Simulations with Proper Force Softening
}

\correspondingauthor{Lehman Garrison}
\email{lgarrison@flatironinstitute.org}

\author[0000-0002-9853-5673]{Lehman H. Garrison}
\affiliation{Center for Computational Astrophysics, Flatiron Institute \\
Simons Foundation, 162 Fifth Ave.\\
New York, NY 10010, USA}

\author{Michael Joyce}
\affiliation{Laboratoire de Physique Nucl\'eaire et de Hautes \'Energies \\
UPMC IN2P3 CNRS UMR 7585 \\
Sorbonne Universit\'e, 4, place Jussieu, 75252 Paris Cedex 05, France}

\author{Daniel J. Eisenstein}
\affiliation{Center for Astrophysics $|$ Harvard \& Smithsonian \\
60 Garden St, Cambridge, MA 02138, USA}



\begin{abstract}
Analysis of self-similarity in scale-free $N$-body simulations reveals the spatial and temporal scales for which statistics measured in cosmological simulations are converged to the physical continuum limit.  We examine how the range of scales in which the two-point correlation function is converged depends on the force softening length and whether it is held constant in comoving or proper coordinates.  We find that a proper softening that reaches roughly 1/30th of the inter-particle spacing by the end of the simulation resolves the same spatial and temporal scales as a comoving softening of the same length while using a third fewer time steps, for a range of scale factors typical to $\Lambda$CDM simulations.  We additionally infer an inherent resolution limit, set by the particle mass and scaling as $a^{-1/2}$, beyond which reducing the softening does not improve the resolution.  We postulate a mapping of these results with spectral index $n=-2$ to $\Lambda$CDM simulations.



\end{abstract}



\section{Introduction} \label{sec:intro}
Detailed comparison of large-scale structure survey data to numerical simulations is a cornerstone of precision cosmology.  In particular, cosmological N-body simulations provide a robust accounting of the gravitational dynamics that govern the clustering of galaxies in a general relativistic framework.  Although non-gravitational physics is ignored in such simulations, they provide a scaffolding for the placement of galaxies with techniques like the halo occupation distribution (HOD) that can marginalize over a plausible range of un-modeled physics (e.g.~\citealt{Zheng+2005}; \citealt{Wechsler_Tinker_2018} for a recent review).

As the resolution of simulations increase, techniques that make use of small-scale structure have ballooned, including sub-halo abundance matching \citep[SHAM,][]{Conroy+2006,Vale_Ostriker_2006} and techniques that use the matter field of the simulations \citep[e.g.~GRAND-HOD;][]{Yuan+2018} rather than assuming dynamical equilibrium and a profile like NFW \citep{Navarro+1997} or Einasto \citep{Einasto_1965}.  Modern galaxy surveys resolve increasingly faint, less massive objects, residing in smaller halos.  In such a context, the convergence of the $N$-body simulation must be established with respect to discretization parameters like the number of particles $N$, the associated mean inter-particle spacing $L/N^{1/3}$ for a given box size $L$, and softening length $\epsilon$ that regularizes the small-scale force.

The precision of simulations matters, too: emulators and other techniques that rely on the derivatives of observables with respect to cosmology require \textit{relative} accuracy.  If the numerical errors in $N$-body simulations are cosmology-dependent, then such derivatives will not be faithfully reproduced.  There are hints of such cosmology dependence, as seen in the redshift dependence of errors in the Euclid code comparison project, for example \citep{Schneider+2016}.

In principle, the convergence of $N$-body simulations to the relevant physical limit---the Vlasov-Poisson solution---can be assessed by extrapolating the discretization parameters appropriately, such as increasing particle count $N$.  In practice, an obstacle to establishing convergence in this way is the computational cost of increasing particle density while also decreasing the force softening length $\epsilon$ \citep[e.g.][]{Power+2003}. Indeed, the the appropriate extrapolation for softening does not send $\epsilon$ to zero while holding the particle spacing fixed, as this amplifies two-body scattering.  Instead, $\epsilon$ must be taken small compared to the physical scales of interest, but still large compared to the particle spacing. Such a regime is, in practice, numerically inaccessible.


Comparison of the $N$-body simulation to a reference solution is therefore desirable but elusive in practice.  One class of test that circumvents some of these difficulties is analysis of self-similarity of scale-free simulations \citep[e.g.][]{Efstathiou+1988,Colombi+1996,Jain+1998,Smith+2003,Widrow+2009,Orban_Weinberg_2011}.  Scale-free simulations are initialized with a power-law power spectrum in an $\Omega_M=1$ (EdS) cosmology, such that there is only one physical scale in the box---the scale of the onset of non-linearity.  The clustering of matter on small scales at early times should be a rescaling of the clustering on large scales at late times, else the simulation is said to break self-similarity.

All simulations will break self-similarity at some level, due to UV cutoffs (finite $N$ or softening), IR cutoffs (finite $L$), or other preferred scales imprinted by the method (e.g.~cell size in the force solver).  But tracking these deviations from self-similarity, and validating that a simulation method has not imprinted any additional scales on the clustering, provides bounds on the range of masses, times, and scales that may possibly be trusted from a simulation.

With this in mind, we have begun a program of exploring the resolution and convergence of scale-free simulations and their data products.  In \cite{Joyce+2020}, we initiated this program with an analysis of the two-point correlation function (2PCF) of matter in a single simulation, employing an $n=-2$ spectral index and a comoving softening of 1/30th of the interparticle spacing.  This work used self-similarity to infer the range of spatial and temporal scales resolved and showed the resolution---the smallest resolved scales---propagates to smaller scales as $a^{-1/2}$ at late times. Further, it concluded that these resolution limits at small scales are set essentially by the mass resolution and should be insensitive to softening provided it is sufficiently small.

In this work, we vary the softening to identify directly its role in setting the small-scale resolution cut-off, and in particular disentangle its contribution from that of finite mass resolution.  We test several different softening lengths, fixed in comoving coordinates, and several fixed in proper coordinates.

Of course, \LCDM simulations are not power-law, EdS cosmologies, and contain many scales besides the onset of non-linearity.  But the non-linear scale is present in both, and does provide a route to creating a mapping from scale-free simulations to \LCDM simulations, as in \cite{Smith+2003} or \cite{Joyce+2020}.  We discuss this below briefly and will explore it more fully in future work.

Self-similarity in a scale-free test does not necessarily ``prove'' that a method is correct or free of errors.  There exist classes of systematic errors that are themselves scale-free---for example, a global time step of any fixed logarithmic size, even implausibly large, will exhibit self-similarity.  However, the primary method we are testing in this work---force softening---manifestly breaks self-similarity.  We thus have a strong theoretical expectation that we should be able to identify scales where self-similarity is broken, and that scales where it is preserved are converged with respect to force softening.


The plan of this paper is as follows.  In Section \ref{sec:softening}, we lay out some theoretical groundwork on $N$-body force softening.  In Section \ref{sec:scalefree}, we introduce the scale-free units in which we will analyze the simulations and define the simulation parameters---a set of identical simulations except for the force softening length and technique.  In Section \ref{sec:results}, we identify the length scales and epochs for which the simulations exhibit self-similarity, and consider how these scales change with softening.  We summarize in Section \ref{sec:summary}.

\section{Softening}\label{sec:softening}
Force softening in $N$-body simulations regularizes the small-scale gravitational force to mitigate the effects of two-body scattering.  To the extent that $N$-body simulations attempt to model collisionless Vlasov-Poisson dynamics, collisional encounters are non-physical and should be suppressed.  However, softening also modifies the small-scale growth of structure, so it should be kept much smaller than the physical scales of interest.  The choice of softening length is thus a balance between amplifying unphysical effects at smaller scales and ``washing out'' clustering on large scales.  A smaller softening length additionally imposes the computational cost of integrating tighter orbits, requiring more time steps.

Choice of softening length has been studied extensively \citep[e.g.][]{Knebe+2000,Power+2003,Diemand+2004,Romeo+2008,Power+2016,Mansfield_Avestruz_2021}.  In principle, the proper procedure to select a softening length is to choose the halo mass of interest, select a softening a few times smaller than the expected core radius of such objects, and then choose $N$ to densely sample the softening length; i.e.~$\epsilon \gg L/N^{1/3}$.  Such a procedure results in smooth, well-sampled dynamics, with limited two-body relaxation and the UV cutoff set only by the softening length.  Of course, increasing $N$ to such extremes is computationally intractable for many science cases.  Within \LCDM halos, one still arrives at the densely-sampled regime in halo cores, but the small softening length increases the importance of two-body scattering.

The effect of softening is primarily determined by its length, although there is also a choice of functional form, of which there are many.  Different forms balance computational expense, compactness (switching to $1/r^2$ at a finite radius), correspondence to some analytic density profile, and avoidance of dynamical instabilities.  For cosmological $N$-body simulations, the softening length seems to matter more than the functional form, except to note that non-compact softening laws can extend the suppression of growth to scales many times larger than $\epsilon$ \citep{Garrison+2019}.

In Plummer softening \citep{Plummer_1911}, the $\bfF(\bfr) = \bfr/r^3$ force law is modified as
\begin{equation}\label{eqn:plummer}
\bfF(\bfr) = \frac{\bfr}{(r^2 + \epsilon_p^2)^{3/2}},
\end{equation}
where $\epsilon_p$ is the softening length.  This softening is very fast to compute but is not compact, meaning it never explicitly switches to the exact $r^{-2}$ form at any radius.

Therefore, many $N$-body codes, such as \textsc{gadget} \citep{Springel+2020} and \Abacus (Garrison et al., in prep; \citealt{Garrison+2019} for a recent description), adopt a spline softening. \Abacus uses the following spline softening from \cite{Garrison+2016}, which switches to the exact form past radius $\epsilon_s$:
\begin{equation}\label{eqn:spline}
\bfF(\bfr) =
\begin{cases}
\left[10 - 15(r/\epsilon_s) + 6(r/\epsilon_s)^2\right]\bfr/\epsilon_s^3, & r < \epsilon_s; \\
\bfr/r^3, & r \ge \epsilon_s.
\end{cases}
\end{equation}

The softening scales $\epsilon_s$ and $\epsilon_p$ imply different minimum dynamical times (an important property, as this sets the time step necessary to resolve orbits).  We always choose the softening length as if it were a Plummer softening and then internally convert to a softening length that gives the same minimum pairwise dynamical time for the chosen softening method.  For our spline, the conversion is $\epsilon_s = 2.16\epsilon_p$.  Under this definition, our spline matches $1/r^2$ more closely than Plummer down to about $0.5\epsilon_p$, below which the two match to within 10\%.

While the softening length is usually taken to be fixed in comoving coordinates, it is sometimes also fixed in proper coordinates, especially in hydrodynamical simulations where galactic structures are effectively ``frozen out'' of cosmic expansion.  Formally, $N$-body leapfrog integration is usually defined in terms of comoving coordinates and canonical momentum, such that any time-varying softening formally breaks the symplectic nature of the integration \citep{Quinn+1997}.  However, the procedure of choosing a time step is often non-commutative with the kick and drift operators and therefore breaks the symplectic property anyway.  Time stepping schemes are an active area of work in $N$-body \citep[e.g., recently,][]{Springel+2020,Zhu_2021}.  For practical purposes, in cosmological simulations, the computational economy of adaptive time stepping is often preferred over the formal advantage of symplecticity.

\begin{figure}
    \centering
    \includegraphics[width=\columnwidth]{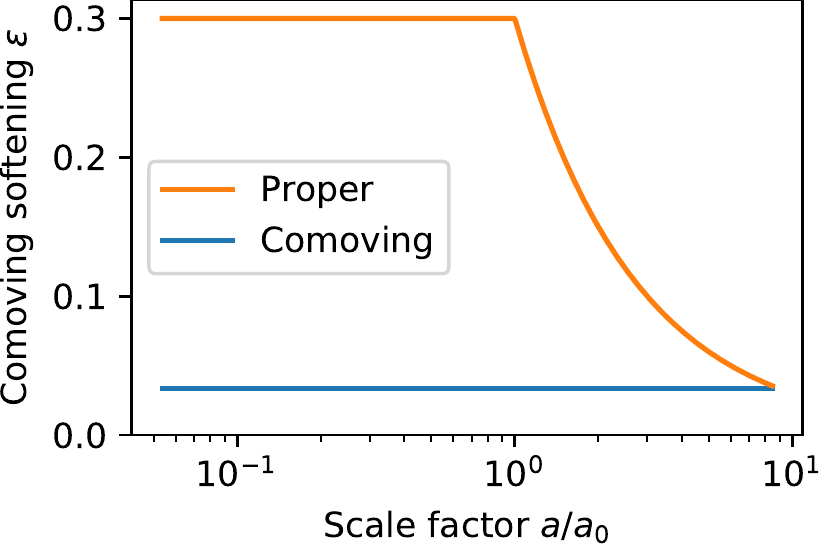}
    \caption{Softening length versus scale factor for the fiducial simulations with comoving and proper softenings.  The proper softening length has a greater comoving value for all epochs considered (Eq.~\ref{eqn:softening_cap} with $\epsilon_\mathrm{prop}=0.3$).
    \label{fig:softening_hist}}
\end{figure}

We test both comoving and proper softening in this work.  Comoving softening is constant in time, since the simulation uses comoving coordinates.  Proper softening increases in comoving coordinates towards earlier times.  Since cosmological simulations typically run for a factor of 10 to 100 in scale factor, proper softening can become inappropriately large at early times if unchecked (e.g.~a softening of $1/30$th of the mean particle spacing at $z=0$ will span three particles in the initial configuration at $z=99$).  This will suppress linear growth of structure beyond what already happens due to the discretization of the mass field \citep{Garrison+2016}.  Furthermore, it will introduce transients, because the Lagrangian perturbation theory assumes a continuum of mass elements that evolve without softening; the introduction of softening at the start of the simulation changes the growth rate discontinuously (unless one solves the Lagrangian perturbation theory with direct force evaluations, as in \citealt{Garrison+2016}).

Therefore, we cap the proper softening at high redshift as
\begin{equation}\label{eqn:softening_cap}
    \epsilon(a) = \min\left(\frac{a_0\epsilon_\mathrm{prop}}{a}, 0.3\right).
\end{equation}
$\epsilon$ is the comoving epsilon employed by the simulation at scale factor $a$, for a given proper softening $\epsilon_\mathrm{prop}$ defined at $a_0$.  This is shown in Figure~\ref{fig:softening_hist} for $\epsilon_\mathrm{prop} = 0.3$.
Effectively, at early times the softening is comoving and fixed to 0.3 in units where the particle spacing is unity, so the worst of the growth rate modifications should be suppressed as it is smaller than the mean interparticle spacing.  Then, at some intermediate redshift, when the proper softening is smaller than 0.3, the softening begins shrinking in inverse proportion to the scale factor.  Note that with this definition, larger $\epsilon_\mathrm{prop}$ does not change comoving plateau value, but only delays the transition from comoving to proper evolution.

Fixing the softening in proper coordinates, $\epsilon_\mathrm{prop} = a \epsilon_\mathrm{com}$, has the effect of increasing the comoving softening length at earlier times.  One should expect the usual effects of increasing softening: suppressed growth of structure on small scales, but decreased relaxation from impulsive encounters.  


In a simulation with only one particle species, mass segregation from two-body relaxation is not a concern.  However, impulsive encounters may still inject kinetic energy into a fraction of particles, stealing potential energy from the remaining particles, leading to core collapse and/or evaporation.  In practice, the time scale of core collapse is tens to hundreds of relaxation times and is unlikely to occur on cosmological time scales; evaporation takes even longer \citep{Knebe+2000,Binney_Knebe_2002,Binney_Tremaine_2008}.  But the population of particles with high kinetic energy is a generic feature that may still be present, perhaps manifesting as ``puffier'' halos.

\section{Scale-Free Simulations}\label{sec:scalefree}
\subsection{Definitions}
In scale-free $N$-body simulations, the particles are initialized with a power-law power spectrum of index $n$ in an $\Omega_M = 1$ background cosmology.  The linear power spectrum grows as:
\begin{equation}
P_L(k,a) = A a^2 k^n
\end{equation}
for scale factor $a$ and amplitude $A$.

The only physical scale in the problem is the non-linear scale, defined by the onset of large-amplitude matter clustering.  In configuration space, we may define this using $\sigma^2(R,a)$, the variance of density in spheres of radius $R$:
\begin{align}
\sigma^2(R,a) &= \int \frac{d^3k}{(2\pi)^3}\tilde W_T^2(kR)P_L(k,a) \nonumber \\
            &= \frac{9 A a^2 R^{-(3+n)}}{2 \pi^{3/2}} \frac{\Gamma[(3+n)/2]}{\Gamma[(2-n)/2] (n-3)(n-1)},
            \label{eqn:sigma}
\end{align}
where $\tilde W_T^2$ is the Fourier transform of the spherical tophat window function.  Eq.~\ref{eqn:sigma} is written in a numerically stable form that avoids divergence of the $\Gamma$ functions for integer $n \in (-3,1)$ \citep{Garrison_thesis}.

We may define $R_\mathrm{nl}$ as the scale at which the dimensionless variance reaches unity, using Eq.~\ref{eqn:sigma} to find a self-similarity relation:
\begin{align}\label{eqn:map}
    \sigma^2(R_\mathrm{nl},a) &= 1 \nonumber \\
    a^2 R_\mathrm{nl}^{-(3+n)} &\propto 1 \nonumber \\
    R_\mathrm{nl} &\propto a^{2/(3+n)}.
\end{align}
Eq.~\ref{eqn:map} defines a map of redshift to scale.  In other words, any function of scale, such as the two-point correlation function, must be a rescaling of the same function at a different epoch.  Indeed, any dimensionless function of scale $r/R_\mathrm{nl}$ must be constant so long as self-similarity holds.

\subsection{Abacus $N$-body Code}\label{sec:abacus}
We employ the \Abacus $N$-body code (described in \citealt{Garrison+2019}; code paper in prep.) for the simulations in this work.  \Abacus is a high-performance, GPU-based code that uses an analytic separation of the near- and far-field forces to solve the former with exact summation and the latter with a high-order multipole method ($p=8$ in this work).  The result is highly accurate forces, such that the remaining error is primarily the finite time step size of the leapfrog integration \citep{Garrison+2019}.  In \cite{Joyce+2020}, we demonstrated that a time step parameter of $\eta_\mathrm{acc}=0.15$ is sufficient for the 2PCF of matter to be converged to 0.1\% for an $n=-2$ simulation.  We confirm that finding in this work and augment it with tests using proper softening.

\Abacus employs a global leapfrog time step, shared by all particles.  The size of the time step, $\Delta a$, is chosen at the beginning of each time step as the greater of a cell-based value, $\Delta a_c$, and a global value, $\Delta a_g$, scaled by the time step parameter $\eta$:
\begin{equation}
    \Delta a = \eta \max(\Delta a_c, \Delta a_g).
\end{equation}
Cells are the unit of the \Abacus cubic grid domain decomposition, with typically a few dozen particles per cell on average. $\Delta a_c$ is computed as the smallest $v_\mathrm{rms}/a_\mathrm{max}$ among cells:
\begin{equation}
    \Delta a_\mathrm{c} = \min_{c'} \left[ \left (\frac{1}{N_{c'}}\sum_i^{N_{c'}} |\mathbf{v}_i|^2 \right)^{1/2} \bigg/ \max_{i<N_c,j<3} a_{i,j} \right],
\end{equation}
where $N_c$ is the number of particles in cell $c$, $\mathbf{v}_i$ is the velocity vector of particle $i$, and $a_{i,j}$ is component $j$ of the acceleration vector of particle $i$.

$\Delta a_g$ is computed analogously, but using the global $v_\mathrm{rms}$ and $a_\mathrm{max}$:
\begin{equation}
    \Delta a_\mathrm{g} = \left[ \left (\frac{1}{N}\sum_i^N |\mathbf{v}_i|^2 \right)^{1/2} \bigg/ \max_{i<N,j<3} a_{i,j} \right],
\end{equation}

The comparison of the cell-based value to the global value guards against abnormally cold cells causing catastrophically small time steps.  Since the velocity RMS is not centered on the center-of-mass velocity, $v_\mathrm{rms}$ in unlikely to be low even in regions of low velocity dispersion, as long as cosmological flows are present.  This may not be the case for all cells, though, especially at late times where the large volume occupied by voids can lead to many cells with low occupation, and therefore a greater chance to find a cold cell.  The global criterion mitigates this.

Additionally, the time step is capped to 0.03 in units of $d\ln a$, ensuring at least 33 steps per $e$-fold of the scale factor.  This criterion dominates at early times, before clustering has brought about small dynamical times in the centers of halos.  This ensures that the accuracy of early linear-theory growth.

\subsection{Simulation Parameters}
The fiducial simulation is a scale-free realization of a $n=-2$, $\Omega_M=1$ cosmology with $N=1024^3$ particles.  We work in units of the mean inter-particle spacing, or $L/N^{1/3} = 1$.

The first output epoch, $a_0$, is chosen based on the top-hat density standard deviation at the particle spacing:
\begin{gather}\label{eqn:a0}
    \sigma(R=1,a_0) = 0.56.
\end{gather}
The value of $0.56=1.68/3$ is chosen so that 3-sigma objects will be reaching the spherical collapse threshold of 1.68.  Subsequent output epochs are spaced by a factor of $\sqrt{2}$ in the non-linear mass scale:
\begin{equation}
    \Delta \log_2 M_\mathrm{nl} = 0.5.
\end{equation}
$M_\mathrm{nl}$ may be defined proportionally to the non-linear length scale $R_\mathrm{nl}$:
\begin{equation}
    M_\mathrm{nl} \propto R_\mathrm{nl}^3.
\end{equation}
From this and Eq.~\ref{eqn:map}, we may determine the spacing of our outputs in scale-factor:
\begin{equation}
\begin{split}
    \Delta \log_2 a_\mathrm{nl} &= \frac{d\ln a}{d\ln M} \Delta \log_2 M_\mathrm{nl}, \\
     &= \frac{3+n}{6} \Delta \log_2 M_\mathrm{nl}, \\
    &= 1/12.
\end{split}
\end{equation}
In the last line, we used the index $n=-2$.  We produce 38 outputs, or about a factor of $5\times10^5$ in $M_\mathrm{nl}$ and 80 in $R_\mathrm{nl}$.

The simulations use 2LPT initial conditions with the configuration-space method of \cite{Garrison+2016}.  The initial scale factor, $a_i$, is chosen so that the density fluctuations are small compared to unity:
\begin{equation}
    \sigma(R=1,a_i) = 0.03.
\end{equation}

We employ particle linear theory (PLT) corrections in the initial conditions to null out transients that arise from the particle discretization of the matter density field \citep{Joyce_Marcos_2007,Garrison+2016}.  On small scales, we apply a wavevector-dependent rescaling of the amplitude of the power spectrum such that the simulations arrive at the correct linear theory value at $a_\mathrm{PLT}$, even though the growth rate remains unavoidably modified.  We choose $a_\mathrm{PLT} = a_0$; we have tested variations in this choice, including disabling rescaling entirely, which generally worsen self-similarity but do not affect our qualitative conclusions.

This work focuses on variations in softening and time step, which are summarized in Table~\ref{table:sims}.

\begin{deluxetable}{rlrrl}
\tablecaption{Simulations used in this work.  For proper softening, $\epsilon(a_0)$ corresponds to $\epsilon_\mathrm{prop}$ in Eq.~\ref{eqn:softening_cap} (with $a_0$ defined by Eq.~\ref{eqn:a0}).
\label{table:sims}}
\tablehead{\colhead{$N$} & \colhead{Softening} & \colhead{$\epsilon(a_0)$} & \colhead{$\eta$} & \colhead{Comment}}
\startdata
$1024^3$ & Comoving & $1/30$ & 0.15 & Fiducial comoving \\
& & & 0.075, 0.3 & Time step variations \\
& & $1/60$, $1/15$ & & Softening variations \\
\hline
$1024^3$ & Proper & 0.3 & 0.15 & Fiducial proper \\
& & & 0.075, 0.3 & Time step variations \\
& & 0.1, 0.5, 0.9 & & Softening variations \\
\enddata
\end{deluxetable}

\subsection{Softening Prescriptions}
All softening in this work uses \Abacus spline softening, defined in Equation~\ref{eqn:spline}.  As discussed in Section~\ref{sec:softening}, proper softening can become inappropriately large at early times, spanning multiple particle spacings and thereby modifying the linear growth of structure.  Therefore, the softening is capped at early times to 0.3, as detailed in Equation~\ref{eqn:softening_cap}.

In our fiducial simulations the proper softening is greater than the comoving softening for all epochs considered, starting $9\times$ greater for $a<a_0$ and ending at an almost identical value. This increases the smallest dynamical times in the simulation, requiring the simulation to take fewer time steps---an advantage in computational economy.  In Section \ref{sec:runtime}, we will discuss that the fiducial proper simulation takes 65\% as many time steps as the fiducial comoving.

Because we are employing compact spline softening, there is an exact switchover radius from softened force to $1/r^2$ (see the discussion in Sec.~\ref{sec:softening}).  For a comoving softening of 0.3 (the maximum value allowed by Eqn.~\ref{eqn:softening_cap}), that switchover radius is about 0.65.  It is not 0.3 because all our softening lengths are quoted as effective Plummer values to facilitate comparison between different softening schemes.  Since this value is less than 1, we expect that the most serious effects of softening modifying the initial growth rate discontinuously from that assumed by the initial conditions generator will be small.

\subsection{Analysis}
As in \cite{Joyce+2020}, our analysis methodology is to evaluate the matter 2PCF of the simulations at all epochs, apply the self-similar rescaling to cast the comparison in units of $R_\mathrm{nl}$, and observe how quickly the simulations converge to the self-similar solution for different length scales.  This is presented in Section \ref{sec:results}.

We use the \textsc{Corrfunc} code of \cite{Sinha_Garrison_2019,Sinha_Garrison_2020} to compute the correlation functions.  The radial bins are scaled self-similarly for each epoch, so no interpolation or centering corrections are required to compare measurements across epochs.  Particles are down-sampled by a factor of 2 before pair counting, and late-time, large-scale bins are elided.  These bins are not needed to assess small-scale convergence and indeed begin to exhibit finite box-size effects, even at 1/100th of the box scale, due to the very red $n=-2$ power spectrum.

\section{Results}\label{sec:results}
\subsection{Correlation Functions}
\begin{figure}[t]
    \centering
    \includegraphics[width=\columnwidth]{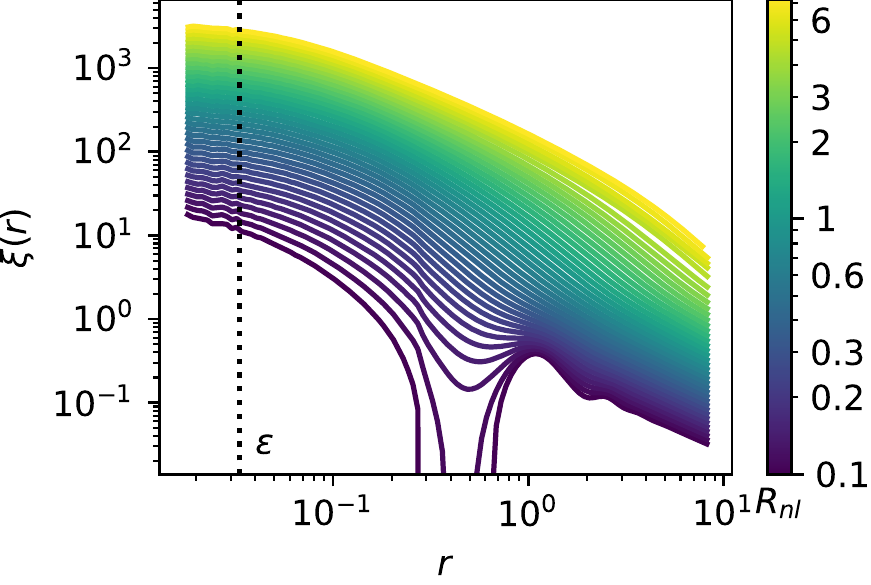}
    \caption{Correlation functions from 38 epochs of the fiducial simulation with comoving softening.  Line color corresponds to non-linear scale $R_\mathrm{nl}$, a label of the epoch, with lighter color indicating later epoch.  Oscillations from the near-lattice structure of the initial conditions are evident at early times near $r=1$ (the particle spacing scale).  The comoving softening length, $\epsilon=1/30$, is marked.}
    \label{fig:raw_comoving}
\end{figure}

\begin{figure}[t]
    \centering
    \includegraphics[width=\columnwidth]{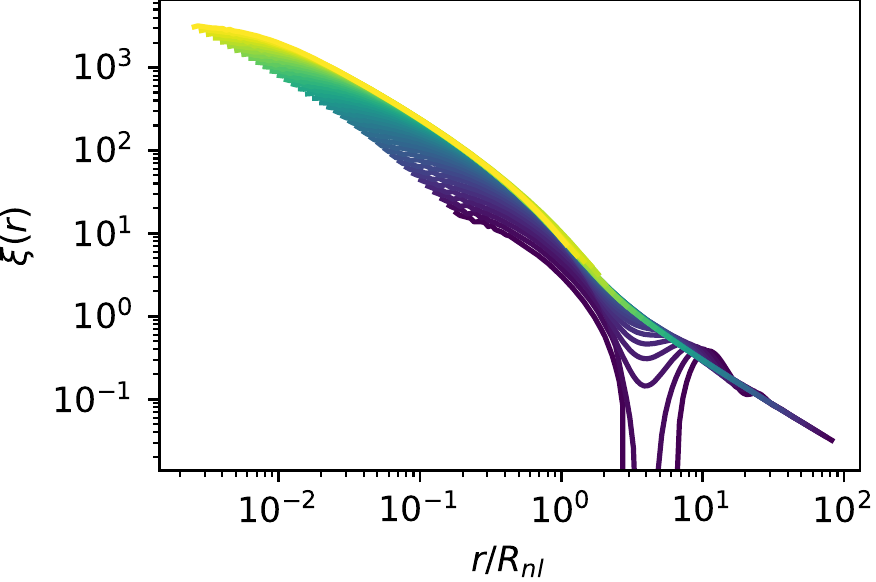}
    \caption{The self-similar rescaling of Figure \ref{fig:raw_comoving}.  The lines lie atop one another, except where self-similarity is broken.  On the left, the ``fanning'' is due to softening and finite mass resolution, and the oscillations from the initial particle lattice are still seen at early times.  By taking vertical slices through this plot, we may assess the rate of convergence as a function of epoch (Figure \ref{fig:convergence_comoving_proper} and onwards).}
    \label{fig:rescaled_comoving}
\end{figure}

\begin{figure}[t]
    \centering
    \includegraphics[width=\columnwidth]{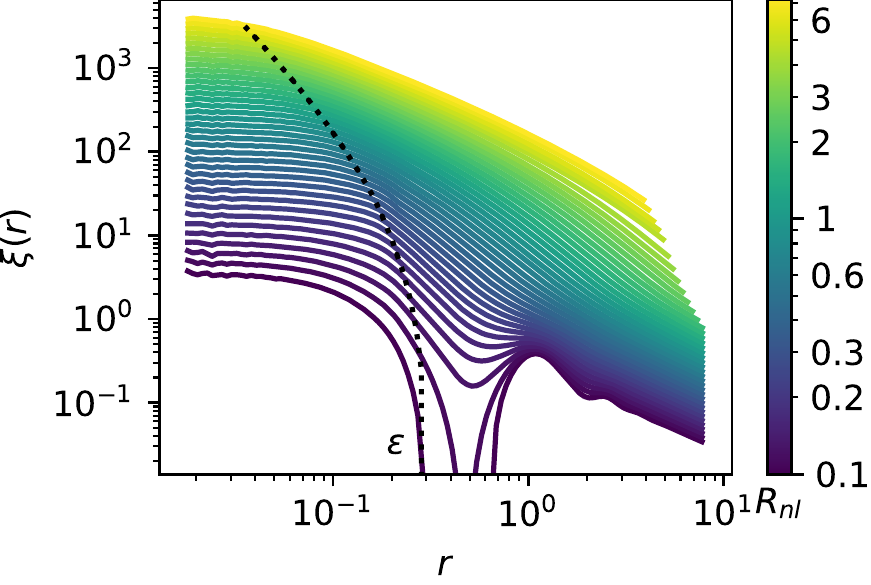}
    \caption{Same as Figure \ref{fig:raw_comoving}, but for the fiducial simulation with softening held constant in proper coordinates.  The comoving softening therefore shrinks towards later epochs, as shown by the dashed line.}
    \label{fig:raw_proper}
\end{figure}

\begin{figure}[t]
    \centering
    \includegraphics[width=\columnwidth]{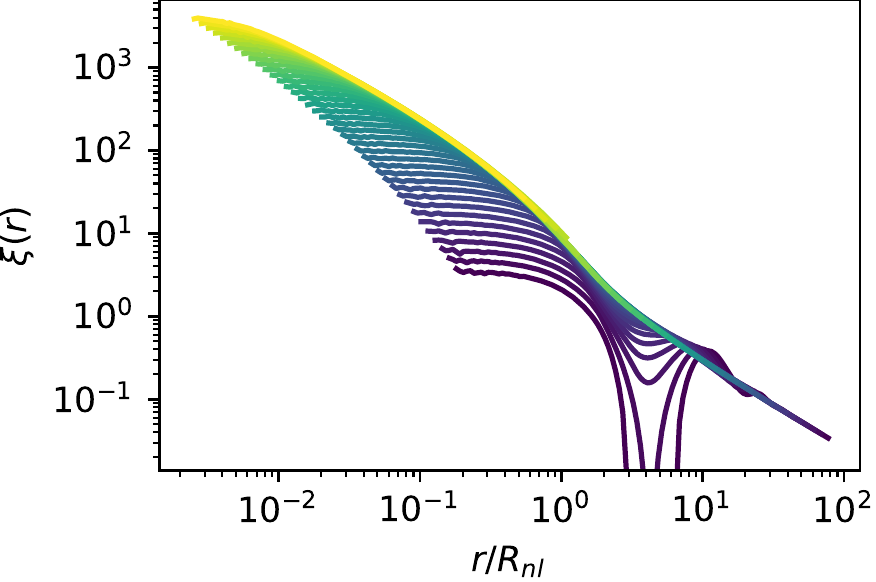}
    \caption{The self-similar rescaling of Figure \ref{fig:raw_proper}.}
    \label{fig:rescaled_proper}
\end{figure}

Figures \ref{fig:raw_comoving} \& \ref{fig:raw_proper} show the raw 2PCF measurements for the fiducial simulations with comoving and proper softening, respectively.  Figures \ref{fig:rescaled_comoving} \& \ref{fig:rescaled_proper} show their self-similar rescalings, using Eq.~\ref{eqn:map}.

First, we see that the rescaled correlation functions largely exhibit self-similarity---that is, they stack.  But there are prominent deviations demonstrating that self-similarity is broken.  On small scales, we see ``fanning'', or flattening, due to a combination of softening and finite particle mass.  At somewhat larger scales, we see dips and oscillations, more clearly seen in the raw correlation functions.  There, we observe that these early-time oscillations fall around $r=1$, and are thus the ``memory'' of the near-lattice structure of the initial conditions (the particle spacing is unity).  This feature is washed out after a factor of a few in $R_\mathrm{nl}$ as the simulation Poisson-izes.

To more precisely assess how quickly different scales converge, we take vertical slices through these rescaled figures and plot each slice as a panel of, e.g., Figure \ref{fig:convergence_comoving_proper} and subsequent figures.  Flat lines---constant correlation amplitude as a function of epoch---indicate convergence to self-similarity.  The range of epochs and $r/R_\mathrm{nl}$ scales that exhibit this convergence are what we refer to as ``resolved scales''.  The impact of variations in softening on resolved scales will be discussed in Section \ref{sec:results}.

A few small, sharp wiggles are visible on the smallest scales of the 2PCF measurements, e.g.~between the minimum $r$ and the softening length $\epsilon$ in Fig.~\ref{fig:raw_comoving}.  This is due to the lossy compression applied to the particle data that truncates the precision of the positions to 14 bits (as cell offsets, so 22.7 bits of global position).

\subsection{Comoving vs.~Proper Softening}\label{sec:comoving_vs_proper}
\begin{figure*}
    \centering
    \includegraphics[width=\textwidth]{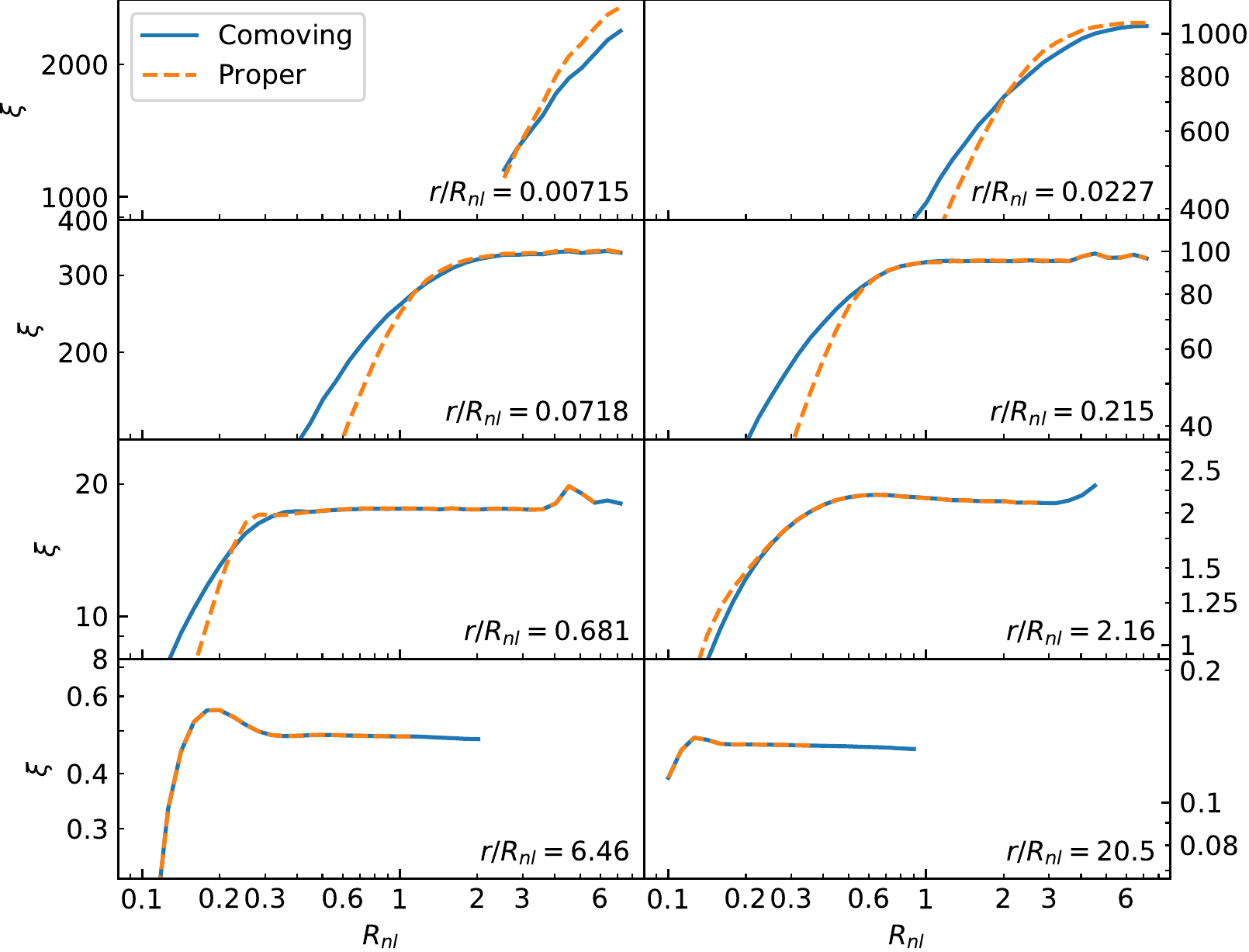}
    \caption{The self-similar convergence of the fiducial simulation for both comoving and proper softening.  Each panel represents a vertical slice in $r/R_\mathrm{nl}$ (see Fig.~\ref{fig:rescaled_comoving}), with flat lines indicating self similarity.  The range of resolved scales is remarkably similar in both, but with proper softening exhibiting a sharper rolloff at early times.  Each panel shows 0.5 dex in $\xi$.}
    \label{fig:convergence_comoving_proper}
\end{figure*}

In Figure \ref{fig:convergence_comoving_proper}, we compare the self-similarity of the fiducial simulation with comoving softening ($\epsilon=1/30$) to that with proper softening ($\epsilon(a_0)=0.3$).  We consider several different values of $r/R_\mathrm{nl}$ (different panels), which, when there is convergence to self-similarity, correspond to different $\xi$ amplitudes.

First, we see that the overall self-similarity---the range of plateaued $\xi$ values---is strikingly similar between comoving and proper.  Indeed, there is even mild evidence that the proper softening increases the range of resolved epochs (e.g.~second panel).  However, at early epochs (small $R_\mathrm{nl}$), the clustering with proper softening falls off more steeply.  This is because the proper softening is largest at early times, and the small scale measurements fall well below the proper softening length as a result. For example, at $R_\mathrm{nl}=1$ in the second panel, $r/R_\mathrm{nl}\sim 0.02$ yields $r\sim 0.02$, which is seen in Fig.~\ref{fig:raw_proper} to fall an order of magnitude below the softening length at early times.  In Fig.~\ref{fig:raw_comoving}, showing the comoving results, $r\sim0.02$ is much closer to the softening.

Despite the proper softening being larger than the comoving softening for the entire duration of the simulation (Figure \ref{fig:softening_hist})---starting from $9\times$ larger at initial output $a_0$---the proper softening arrives at a more clustered state on small scales at late times (e.g.~first panel).  This is somewhat surprising, as considering the growth of modes, a larger softening decreases the growth rate.  But considering halo dynamics, it is plausible that proper softening decreases two-body relaxation, which is dependent on the amplitude of the impulsive scattering, leading to more concentrated halos.

\begin{figure}
    \centering
    \includegraphics{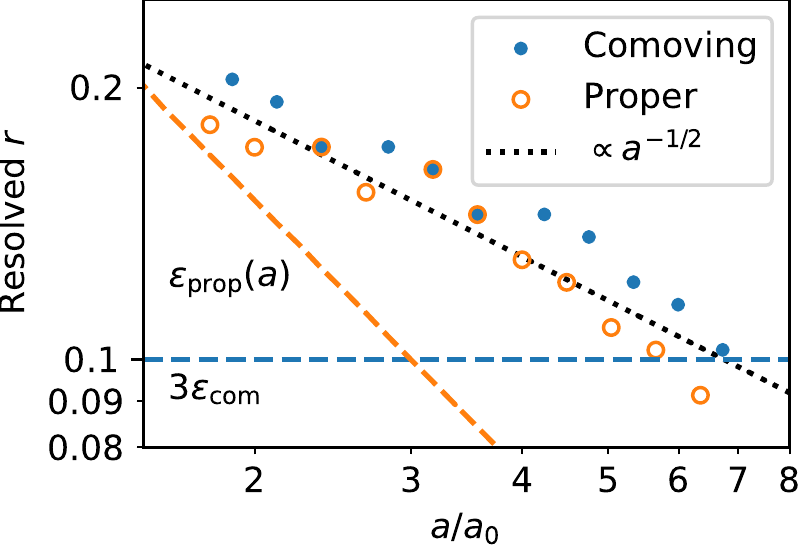}
    \caption{The minimum resolved scale $r$ versus epoch for the fiducial comoving and proper simulations.  $r$ is given in units of the comoving inter-particle spacing, and the epoch is relative to the first output epoch $a_0$ (Eq.~\ref{eqn:a0}).  The comoving simulation has $\epsilon = 1/30$ (blue dashed line shows $3\times$ this value) and the proper $\epsilon(a_0)=0.3$ (orange dashed line).
    Dots and circles are measurements of the epoch at which the correlation function is within 5\% of the converged value (Fig.~\ref{fig:convergence_comoving_proper}).  Overall, they agree very well, despite the proper softening length being larger for all epochs. The black dotted line is a fit by eye of the hypothesis $a^{-1/2}$.}
    \label{fig:com_proper_res}
\end{figure}

We can quantify the range of resolved scales by identifying the epoch $R_\mathrm{nl}$ at which the correlation function first reaches within 5\% of its plateau value.  Repeating this for each slice of $r/R_\mathrm{nl}$ (each panel), we can map the smallest resolved scales versus $a/a_0$.  This is shown in Figure~\ref{fig:com_proper_res}.  The axes are no longer in rescaled units; this analysis connects scale-free units back to comoving units of the inter-particle spacing.

We see that the comoving and proper softenings largely agree which scales are resolved, with some evidence of better convergence (smaller resolved $r$) with the proper softening at early and late times, as observed in Fig.~\ref{fig:convergence_comoving_proper}.  But the measurements are noisy due to binning and finite number of output epochs, so we do not focus on detailed differences---as small as one bin in most cases---but rather broad agreement.

The resolved $r$ propagates to progressively smaller scales as $a^{-1/2}$ to a good approximation.  This supports the hypothesis set forth in \cite{Joyce+2020} that two-body relaxation causes the breakdown of self-similarity on small scales, and the breakdown occurs at a smaller comoving scale as the average number of particles within that scale increases.  In other words, two-body relaxation sets the effective resolution of the simulation.  The $a^{-1/2}$ behavior is consistent with the two-body scattering of clustering that is approximately constant in proper coordinates---so-called ``stable clustering'' \citep{Efstathiou+1988,Colombi+1996,Jain+1998,Widrow+2009}.

The comoving and proper softening lengths are marked by dashed lines in the figure.  The proper softening evolves with time in comoving units, but at all resolved epochs, the resolution limit is above the proper softening.  So although we achieve similar or better results than with comoving softening by employing a proper softening that is larger for all epochs, it is likely that the proper softening must still be smaller than the resolution limit set by the particle mass, lest the simulation be softening-limited.  This is supported by our findings in the next section.

While we have focused on the range of resolved scales, we should also revisit the behavior outside those scales.  As mentioned above, proper softening suppresses the unresolved clustering at early times (although sometimes enhances it at late times).  The suppression effect can be large---a factor of several, as one can see comparing Figures \ref{fig:raw_comoving} \& \ref{fig:raw_proper} by eye.  Where using such scales cannot be avoided, it is already common practice to apply flexible analysis to marginalize over the details of such clustering.  These results confirm the importance of such robust analysis methods and understanding where such unresolved scales begin.

In Fig.~\ref{fig:convergence_comoving_proper}, for large $r/R_\mathrm{nl}$, the comoving and proper softenings agree perfectly, as we would expect for scales well above the UV cutoff.  Deviations from self-similarity start to appear on the largest scales at the latest times, which are most likely finite box size effects.  The length scales at which such effects start to appear, around 1/300th of the box size, are smaller than normal large-box, \LCDM intuition would indicate, but are not surprising given the power-law power spectrum with a red, $n=-2$ index.

\subsection{Softening Length}
\begin{figure*}
    \centering
    \includegraphics[width=\textwidth]{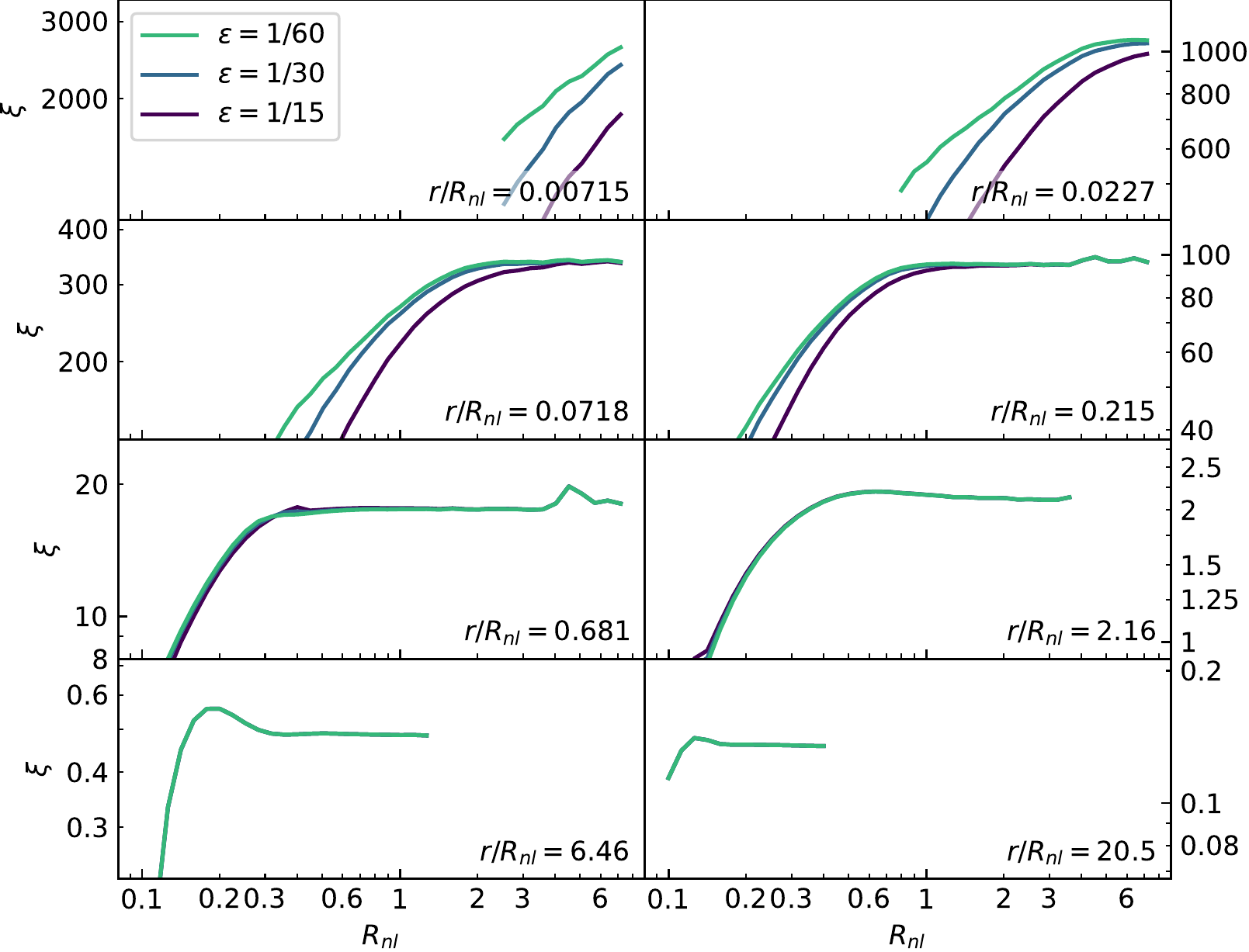}
    \caption{The self-similar convergence with respect to comoving softening length.}
    \label{fig:convergence_comoving_eps}
\end{figure*}

We vary the softening length of the comoving simulation in Figure \ref{fig:convergence_comoving_eps} and the proper simulation in Figure \ref{fig:convergence_proper_eps}.  In the comoving case, we consider $\epsilon=1/60$ and $1/15$, in addition to the fiducial $1/30$.  We find that the range of resolved scales substantially improves from $\epsilon=1/15$ to $1/30$, showing nearly a factor-of-two improvement for large correlation amplitudes (small $r/R_\mathrm{nl}$). Halving the softening again from $1/30$ to $1/60$ shows only marginal improvement; the range of resolved epochs increases only about 10\%.  In unresolved epochs, the clustering amplitude does increase, but again, not as much as the first halving from $1/15$ to $1/30$.

From the decreasing rate of improvement, we infer that there is a resolution limit set by the particle mass beyond which no gain in resolved scales is possible by decreasing the softening length.  This is expected, as the particle mass is the other UV scale besides softening in the problem.   A corollary of this result is that diminishing returns with respect to softening do not indicate a given scale is resolved---this is easily seen in Fig.~\ref{fig:convergence_comoving_eps}, where many epochs are unconverged and yet do not improve with softening length.

We caution that the results of \cite{Joyce+2020} suggest that convergence propagates to progressively smaller scales during the course of the simulation, in proportion to $a^{-1/2}$ at late times.  This result is consistent with a hypothesis that the mass resolution limits the resolved scales through two-body scattering.  Therefore, if we were to run our simulations for longer---to higher clustering amplitudes---the convergence would propagate to small enough length scales that the comoving softening would once again be the limiting factor.  Therefore, the choice of softening length for comoving softening must be made not only with respect to mass resolution but also with respect to the range of scale factors over which the simulation will be run.

Proper softening, which scales as $a^{-1}$, shrinks faster than the resolution, so it does not share this consideration (Fig.~\ref{fig:com_proper_res} shows these scalings). Instead, it must be kept smaller than the resolved scales at early times; e.g.~near $a_0$, or whatever the first epoch of analysis is.  As long as this condition is satisfied, the proper softening will always be smaller than the resolved scales for analysis epochs.

\begin{figure}
    \centering
    \includegraphics{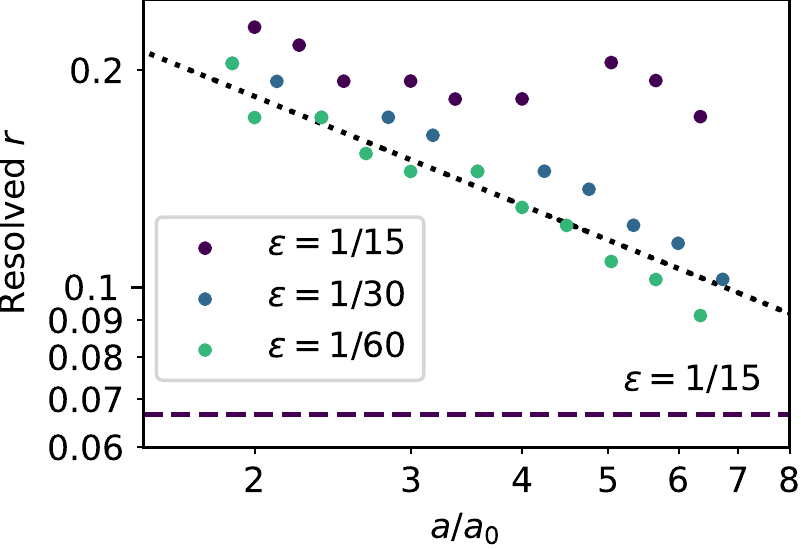}
    \caption{Same as Fig.~\ref{fig:com_proper_res} but for comoving softening only, showing variations in softening length.  As we observe in Fig.~\ref{fig:convergence_comoving_eps}, halving the softening length from 1/15 to 1/30 produces a noticeable gain in resolved scales (smaller $r$) at late epochs where the resolution, nominally improving as $a^{-1/2}$ (black dashed line) approaches the softening.  Halving again to 1/60 produces a much smaller gain; indeed, this gain is only one bin in epoch.  The dark purple dashed line shows the largest of the three softening lengths, $1/15$.}
    \label{fig:com_eps_res}
\end{figure}

Figure~\ref{fig:com_eps_res} quantifies the resolved scales versus epoch for different values of the comoving softening length.  As with Fig.~\ref{fig:com_proper_res} for the comparison of comoving and proper softening, we compare the epochs at which the simulation reaches within 5\% of the plateau value in Fig.~\ref{fig:convergence_comoving_eps} and infer a minimum resolved scale in units of the comoving inter-particle spacing.  We focus on overall behavior rather than detailed differences, because our precision is limited by binning and the spacing of our output epochs.  We see that $\epsilon=1/30$ and $\epsilon=1/60$ agree rather well, with the minimum resolved scale improving by much less than a factor of two---indeed, only by one bin in epoch.  Doubling the softening length from $\epsilon=1/30$ to $\epsilon=1/15$ degrades the resolution by a similarly small amount, except at late epochs where the degradation becomes large.  Indeed, we see nearly the full factor of 2 one would expect if the resolution were determined fully by the softening at these late epochs.  As anticipated in our discussion above, the effect is likely only seen at late epochs because the resolution set by the particles mass propagates to smaller scales as the simulation progresses (hypothesized as $a^{-1/2}$).  When this resolution approaches a few times the softening length, the softening length becomes the limiting factor.  In this case, we observe degradation at about $3\times$ softening length.

\begin{figure*}
    \centering
    \includegraphics[width=\textwidth]{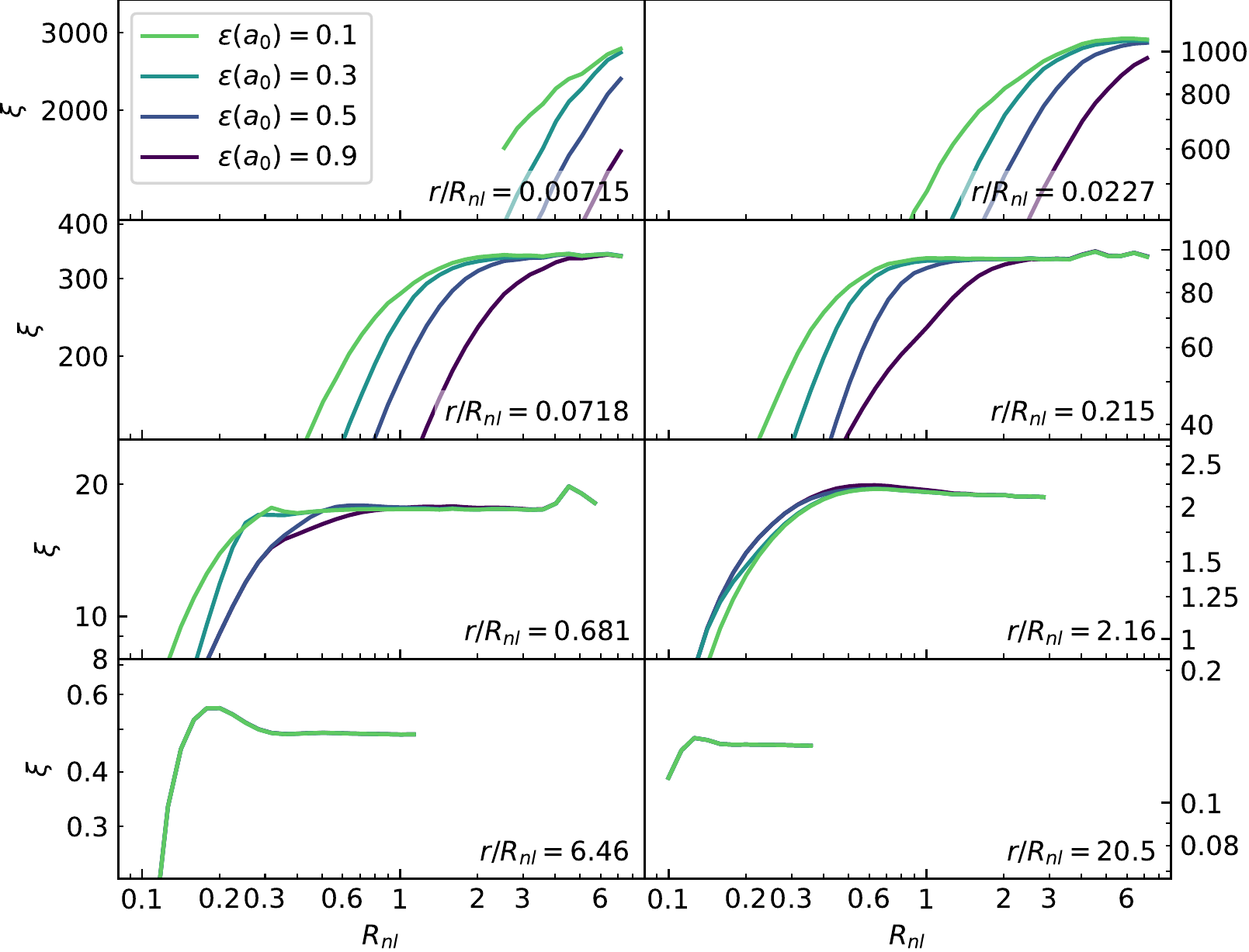}
    \caption{The self-similar convergence with respect to proper softening length.  The softening values $\epsilon(a_0)$ are defined at the epoch $a_0$ of initial output.}
    \label{fig:convergence_proper_eps}
\end{figure*}

We vary the softening of the proper softening simulation in Figure~\ref{fig:convergence_proper_eps}, using values $\epsilon(a_0) = 0.1$, $\epsilon(a_0) = 0.5$, and $\epsilon(a_0) = 0.9$, in addition to the fiducial value of $\epsilon(a_0) = 0.3$.  These epsilon values are given at $a_0$, the epoch of first output, but recall that the resulting comoving value is capped at early times (Eq.~\ref{eqn:softening_cap}).

\begin{figure}
    \centering
    \includegraphics[width=\columnwidth]{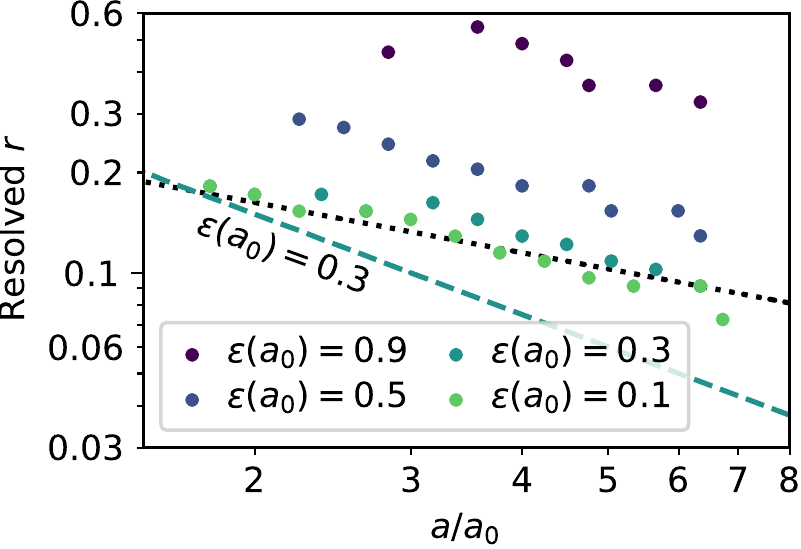}
    \caption{Same as Figs.~\ref{fig:com_proper_res} \& \ref{fig:com_eps_res} but for proper softening only, showing variations in softening length.  As seen in Fig.~\ref{fig:convergence_proper_eps}, taking $\epsilon(a_0)$ from $0.9$ to $0.5$ and $0.5$ to $0.3$ both produce substantial gains in the resolved scale $r$.  Going from $0.3$ to $0.1$ produces a minimal gain---at most one bin of epoch.  The dashed line shows the comoving evolution of the $0.3$ softening, and the dotted line shows the hypothesis $a^{-1/2}$ for the lower resolution limit as set by the particle mass.}
    \label{fig:proper_eps_res}
\end{figure}

Figure~\ref{fig:proper_eps_res} shows the inferred resolution for each softening.  The first factor-of-3 decrease from $\epsilon(a_0) = 0.9$ to $\epsilon(a_0) = 0.3$ makes quite a large difference in the range of resolved epochs---nearly a full factor of 3.  The second factor of 3, from $0.3$ to $0.1$, makes almost no difference---at most one bin of epoch.  The clustering in the unresolved regions does benefit somewhat, however (Fig.~\ref{fig:convergence_proper_eps}).  We infer diminishing returns past $\epsilon(a_0) = 0.3$ and use the intermediate case of $\epsilon(a_0) = 0.5$ to assess this.  We see that there are still moderate gains in decreasing from 0.5 to 0.3, suggesting our fiducial choice of 0.3 essentially reaches the mass resolution limit while not taking too many unnecessary time steps due to an over-small softening.

\subsection{Time Step}
Time step accuracy is difficult to assess with self-similarity analysis \citep[see fig.~2 and discussion in][]{Joyce+2020}.  In particular, if a simulation uses a constant, global leap frog time step in $\log(a)$, then the simulation would be expected to be self-similar for any step size, due to the scale-free nature of logarithmic time stepping, even though the clustering is far from the physical solution.  The \Abacus time step scheme is not quite this simple---using a constant log time step for the earliest times and switching to a dynamical criterion at the onset of clustering (\S\ref{sec:abacus})---but we have observed that self-similarity is substantially upheld even with implausibly large time steps that underestimate the raw clustering amplitude by factors of many.

\begin{figure*}
    \centering
    \includegraphics[width=\textwidth]{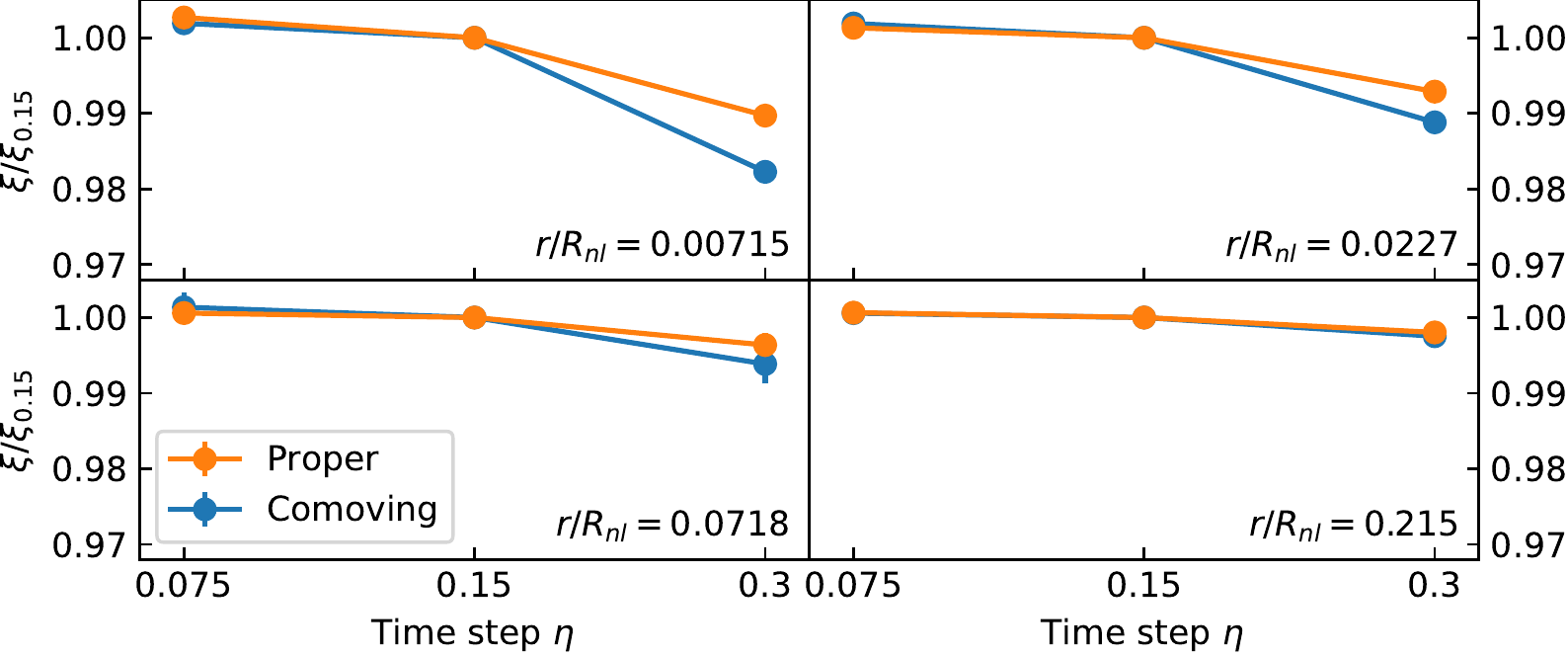}
    \caption{Time step convergence for the fiducial comoving and proper simulations at four different $r/R_\mathrm{nl}$ values (panels).  Ratios are plotted relative to the fiducial simulation with $\eta=0.15$ (i.e.~the middle point).  The behavior is nearly independent of epoch in $r/R_\mathrm{nl}$ units, but we show the mean and standard deviation over epochs for completeness.}
    \label{fig:convergence_comoving_proper_eta}
\end{figure*}

As this suggests, the solution is to directly compare the correlation functions between simulations of different time step, rather than assess a single simulation's self-similarity.  This is shown in Figure \ref{fig:convergence_comoving_proper_eta} for three different time step values: $\eta = 0.3$, $0.15$ (the fiducial), and $0.075$.  We take the fiducial value as our reference not because we expect it to be the most accurate, but because we do not want to suggest that the smallest time step value that we try ($0.075$) is necessarily a converged reference value.  Instead, we focus on the rapidly diminishing change in the $\xi/\xi_{0.15}$ ratio as we halve from $\eta=0.3$ to $\eta=0.15$ and then again to $\eta=0.075$.

The self-similarity of the time stepping error suggests that the ratios of the 2PCF between different simulations may themselves be self-similar.  And indeed this is what we observe.  The time step error is approximately constant as a function of $r/R_\mathrm{nl}$.  Or, perhaps more intuitively, the time step error is only dependent on the value of $\xi$, rather than the physical scale.  Therefore, Figure \ref{fig:convergence_comoving_proper_eta} shows one 2PCF ratio for each simulation for several $r/R_\mathrm{nl}$ values (the same ones considered in the convergence of the softening length).  In detail, there is some small evolution of the 2PCF ratio even in scale-free coordinates, so the mean ratio across epochs is plotted, along with the standard deviation as an error bar.

We see that halving the time step from our fiducial choice in this work, $\eta=0.15$, to 0.075 makes almost no difference, with the largest difference of 0.2\% seen in the smallest scales.  Therefore, we consider our fiducial time step converged for the purposes of this work.

The proper softening simulation is seen to exhibit slightly less dependence on time step than the comoving simulation.  At early times, the proper simulation has likely taken more time steps relative to the dynamical times of collapsing objects because \Abacus's early log-constant time step.  The $\eta=0.3$ simulation takes 103 time steps to the first output, for example, while the $\eta=0.15$ simulation takes 108 time steps---clearly not double.  Therefore, one might posit that the proper softening's relative insensitivity to $\eta$ is because the first 100 steps make up a larger fraction of the time steps.

By the time the last output is reached, the proper simulation has taken 667 steps with $\eta=0.3$ and 1252 with $\eta=0.15$, and the comoving simulation 997 and 1964 steps---ratios of 1.88 and 1.97, respectively.  While the direction of this effect accords with our hypothesis, it is unclear if this small difference can explain the proper simulation's error at $\eta=0.3$ being about 2/3 that of the comoving simulations.

To contextualize these time step parameters, we can consider the number of steps to accomplish the last $e$-fold of the simulation, where the correlation amplitudes reach several thousand.  The fiducial comoving simulation with $\eta=0.15$ takes about 1000 time steps to complete the last $e$-fold, and the fiducial proper about 850.  An AbacusSummit $\Lambda$CDM simulation would take about 1200 steps for the last $e$-fold if it were to use that same time step parameter, with a comoving softening of $1/40$ and particle mass $2\times10^9\ \hMsun$.



\subsection{Run Time}\label{sec:runtime}
To reach the final output, the fiducial simulation took 1964 steps with comoving softening and 1252 steps with proper softening.  This is approximately 65\% as many steps, coming at apparently no cost to the range of resolved scales, but suppressed clustering in unresolved scales, as discussed in Section \ref{sec:comoving_vs_proper}.

\section{Summary}\label{sec:summary}
We have evaluated the self-similarity of scale-free $N$-body simulations with an $n=-2$ spectral index to assess how the range of resolved scales changes with respect to softening length and whether it is held constant in comoving or proper coordinates.  By analyzing the 2PCF and measuring the epoch at which it plateaus to the self-similar value, we have mapped the resolution limit versus scale factor (Figure~\ref{fig:com_proper_res}).  Comparing the resolution limit for a fiducial comoving softening of 1/30 (in units where the inter-particle spacing is 1) with a proper softening that is larger for all epochs, we have found that the proper softening is converged over the same or an even greater range of scales while using 35\% fewer time steps.  Outside of these resolved scales (at earlier epochs or smaller scales), the 2PCF is generally suppressed, although at late times the clustering amplitude is even higher.  We have posited that this increase may be due to the suppression of two-body relaxation.

Considering only comoving softening and varying the softening length, we have found that the minimum resolved scale improves at late times by nearly a factor of two as the softening is halved from $\epsilon=1/15$ to $\epsilon=1/30$.  The next halving, from $\epsilon=1/30$ to $\epsilon=1/60$, shows much less improvement.

We have found, as reported by \cite{Joyce+2020}, that resolution propagates to smaller scales at later times as $a^{-1/2}$.  This is consistent with two-body scattering causing a breakdown of self-similarity on small scales, thereby setting the effective resolution of the simulation.  As the number of particles within a given comoving scale increases, this breakdown moves to smaller scales.  The $a^{-1/2}$ behavior is consistent with the relaxation of halos whose clustering is constant in proper coordinates (stable clustering).  This also explains the advantage of a smaller softening at later times, as produced by proper softening: the softening is not the limiting factor until this resolution, set by the particle mass, approaches the softening scale.  Once this mass resolution limit is reached, we infer there is no gain in resolved scales to be made by taking the softening smaller.

Considering only proper softening and varying the softening length, we similarly have found diminishing returns beyond our fiducial choice.  We note that our choice runs close to the early-time resolution limit set by the particle mass, such that increasing further degrades resolution in this regime.  However, since proper softening shrinks in comoving coordinates faster than the resolution ($a^{-1}$ versus $a^{-1/2}$), the late-time clustering should be less susceptible.

With the mass resolution limit scaling as $a^{-1/2}$, it is natural to consider whether a softening length that scales in the same way would be advantageous.  While this may match the two UV scales well, we do consider proper softening to have a better physical motivation than such a softening.  With proper softening, we suppress secular evolution of halo profiles in proper coordinates due only to the change in softening length.

With respect to time step, we have shown that an \Abacus time step size of $\eta=0.15$ is converged for the accuracy required in this work, and that the time step error is approximately self-similar---in other words, dependent only on the amplitude of the 2PCF, not the epoch.  The proper softening shows relatively less sensitivity to the time step parameter than the comoving softening.

With respect to \LCDM, one might construct a mapping to these $n=-2$ results by identifying $a_0$ for the \LCDM simulation using the epoch of collapse of the first halos (Eq.~\ref{eqn:a0}), as elaborated in \cite{Joyce+2020}.  We will explore the accuracy of such a mapping in future work, including the sensitivity to spectral index.  However, because the small-scale \LCDM index is more negative than $n=-2$, a \LCDM simulation will have more clustering at scales above the radius whose mass variance defines $a_0$.  Since structure is collapsing on larger scales for a given $a_0$---the mass variance is flatter with respect to scale---we expect the sensitivity to softening length to be smaller.  Therefore, we expect that our findings with respect to required softening length are conservative.

The connection of the $a^{-1/2}$ mass resolution limit to two-body relaxation has been derived using scaling arguments in \cite{Joyce+2020}, but closer investigation to establish it could be undertaken with halo-level studies of particle dynamics.  In particular, one could match individual halos between simulations with different softenings and measure the phase-space distribution of particles in the inner cores versus epoch.  We plan to explore this subject in future work.

Aside from spectral index, our analysis was also restricted to the 2PCF.  We will explore other summary statistics and spectral indices in future work.  Such extensions will allow robust extrapolation of these results to the \LCDM simulations that underlie much of the interpretation of large-scale structure data.



\acknowledgments
The authors would like to thank David Weinberg, Yin Li, and the referee for careful reading of the manuscript and helpful comments.  DJE is supported by U.S. Department of Energy grant DE-SC0013718
and as a Simons Foundation Investigator.  Abacus development has been supported by those funds and by NSF AST-1313285 and Harvard University startup funds.  We would like to thank the co-authors of \Abacus: Nina Maksimova, Doug Ferrer, Marc Metchnik, and Philip Pinto.

%


\software{NumPy \citep{numpy},
astropy \citep{Astropy_2018},
Matplotlib \citep{matplotlib}
          }
          
\section*{Data availability}
The correlation function measurements used in this work are available upon request.  The underlying simulation data is substantially larger but may also be made available upon reasonable request.






\bibliography{biblio}{}
\bibliographystyle{aasjournal}



\end{document}